\documentclass{ckm}                 

\usepackage{mathptmx}          
\usepackage{graphicx}
\usepackage{epsfig}

\confname{Workshop on the CKM Unitarity Triangle, IPPP Durham, April
  2003}

\title{Charmed Meson Decays and QCD Sum Rules}

\author{Alexander Khodjamirian \thanks{On leave from
Yerevan Physics Institute, 375036 Yerevan, Armenia}}
\address{Institut f\"ur Theoretische Teilchenphysik, Universit\"at
Karlsruhe, D-76128 Karlsruhe, Germany }

\begin{document}

\begin{abstract}
The current status of the QCD sum rule predictions
for charmed mesons is overviewed.
\end{abstract}

\maketitle


\section{Introduction}

Weak decays of charmed mesons provide a useful testing 
ground for nonperturbative methods, such as QCD sum rules \cite{SVZ}.  
Contrary to $B$ decays, in $D$ decays  the
hadronic matrix elements are multiplied by experimentally 
known CKM parameters: $V_{cs}$ is extracted 
from charm-tagged $W$-decays and $V_{cd}$ from neutrino-nucleon 
production of charm \cite{PDG02}.  
Therefore, the QCD sum rule predictions for the hadronic
parameters of $D$ decays 
can be directly compared with experimental data. 
This comparison allows one to gain more confidence
in the sum rule results for the $B$-decay 
hadronic matrix elements that are used to extract the poorly 
known CKM parameters.

The  outline of the QCD sum rule method and the predictions 
for $B$ decays are discussed in the context of CKM physics
in \cite{CKMyellow2002} 
(in sect.~4.2 of chapter III  and in sects.~2.1.3, 2.1.4 of 
chapter IV), see also \cite{BallCKM}. Here I will concentrate on 
the corresponding results for the charm sector.
A detailed review of QCD sum rules can be found in \cite{CK}.

\section{Determination of the $c$-quark mass}

One of the first applications of  
QCD sum rules was  the estimate of the charmed quark mass.
According to the original method \cite{6auth}, 
one employs the correlator of two $\bar{c}\gamma_\mu c$ 
currents. 
Due to dispersion relation, the n-th derivative 
of the correlator in $q^2$ (the momentum transfer squared)   
is related to the n-th power moment of the
hadronic $e^+e^-\to c\bar{c}$ cross-section $\sigma_c$:
\begin{equation}
M_n(q^2) =\frac{1}{n!}\left(\frac{d}{dq^2}\right)^n \!\!\Pi(q^2) \sim
\!\!\!\!\int\limits_{m_{J/\psi}^2}^{\infty} ds 
\frac{s~\sigma_c(s)}{(s-q^2)^{n+1}}\,.
\label{moments}
\end{equation}
The moments $M_n(q^2)$ are calculated in QCD in a certain range of $\{n,q^2\}$,
far off-shell ($q^2\leq 0$), in terms of the virtual c-quark
loops, taking into account the $c$-quark interactions with perturbative
gluons and nonperturbative gluon condensate \cite{SVZ}. The results 
depend on  $m_c$, $\alpha_s$ and the gluon condensate density.    
The hadronic cross section in Eq.~(\ref{moments})
is saturated by the charmonium resonances $J/\psi,\psi',...$, and at
$\sqrt{s}>2m_D$ by the open charm-anticharm production. 

In Table 1 the recent results  
for the $\overline{MS}$ mass $\bar{m}_c(\bar{m}_c)$
extracted from the sum rule moments (\ref{moments})
with the $O(\alpha_s^2)$ accuracy
are presented and compared with the predictions of other methods.
\begin{center}
\begin{table}[h]
\vspace{-0.6cm}
\caption{}
\begin{tabular}{| c | c | c |}
\hline 
 ${\overline m_c}(\overline m_c)$ (GeV) &Ref.&method\\
\hline
&&\\
$1.304(27) $   &\cite{KuhnStein01}  & QCD SR        \\
$1.275(15)$ &\cite{IoffeZyablyuk02} & ''\\
&&\\
$1.230(90)$ &\cite{EJamin}& QCD SR+NRQCD \\
&&\\
$1.190(110)$ &\cite{Eidemuller}& '' \\
&&\\
\hline
&&\\
$1.370(90)$ &\cite{PenSchilcher01} & FESR \\
&&\\
$1.210(70)(65)(45)$ &\cite{Pineda01} & $m_B-m_D$ + HQET \\
&&\\
\hline 
&&\\
$1.300(40)(200)$&\cite{Lellouch}& lattice QCD (average)\\ 
&&\\
\hline 
\end{tabular}
\end{table}
\end{center}
\vspace{-1.0cm}
Despite a reasonable agreement between all four QCD SR predictions  
within theoretical uncertainties, one has to keep in mind that the 
quoted estimates of $\bar{m}_c(\bar{m}_c)$  are  
obtained using $M_n$ with 
different $\{n,q^2\}$ and making different assumptions. 
In \cite{KuhnStein01}, following the original analysis of
\cite{6auth} the lowest moment $n=1$ at $q^2=0$ 
is selected, having
a little sensitivity to nonperturbative effects 
but demanding an accurate knowledge of the 
cross section $\sigma_c(s)$  above the open charm threshold.
In the analysis of \cite{IoffeZyablyuk02}, $ n\sim 10$  
and $q^2<0$ are used, so that the gluon condensate
contributions become important. 
Finally, in \cite{EJamin,Eidemuller}, for the moments
with large $n$,   an attempt is made to account 
for the resummed Coulomb effects that are not accessible 
in the relativistic calculation of $\Pi(q^2)$.  
An ansatz for the spectral density of the 
correlator is adopted, combining the 
full QCD answer  with the resummed NRQCD spectral density at
large and small $c$-quark velocities, respectively. 
This choice emphasizes Coulomb versus gluon-condensate 
effects in a relatively light  $c\bar{c}$ quarkonium,
an issue which deserves further studies 
(for a critical  discussion see \cite{IoffeZyablyuk02}).  
In order to further improve the sum rule determinations 
of $m_c$, 
one needs more precise measurements of the $J/\psi,\psi',...$ leptonic
widths and of the open charm cross section in $e^+e^-$.

Furthermore, let me remind that the vector charmonium
channel provides a  possibility to check 
the quark-hadron duality approximation, the key element
of QCD sum rules.
Replacing in the integrand  in Eq.~(\ref{moments})
the hadronic cross section by the  calculated spectral density of the 
correlator at $s>s_0$, one still successfully fits the 
moments, even if $s_0$ is shifted close to the 
open charm threshold, indicating  that the ``semi-local''
quark-hadron duality also works in this channel.

\section{$f_D$ from SVZ  Sum Rules } 

One of the great advantages of the sum rule method 
is a possibility to easily switch 
from $c$ to $b$ quark and vice versa, in the analytical
expressions. 
An important example is 
the calculation of the $D$-meson decay constant $f_D$ 
defined as   
$f_Dm_D^2=m_c\langle 0| \bar{d}i\gamma_5 c |D^+\rangle $.  
The SVZ sum rule is derived from the 
correlator of two $\bar{c}i\gamma_5 d $ currents,
applying duality approximation for excited $D$ states
and Borel transformation. The result has the following schematic
form:
\begin{eqnarray}
&&f_D^2m_D^4 e^{-m_D^2/M^2}=\nonumber 
\\
=&&\!\!\!\!\!\!\sum\limits_{n=0,3,...}^{n_{max}}\left( 
\sum\limits_{k=0,1,..} \left(\frac{\alpha_s}{\pi}\right)^k
S_{nk}(m_c, M^2,s_0^D,\mu_c) L_n(\mu_c)\right) 
\label{fDsr}
\end{eqnarray}
where $S_{nk}$ are the calculable short-distance
coefficients. In particular  $S_{0k}$
are given by perturbative heavy-light loops in order $\alpha_s^k$ 
($L_0\equiv 1$); $L_{n\geq 3}$ are the universal long-distance parameters
(vacuum condensate densities) with dimension $n$. In the above, 
$M^2$ $\sim m_c\overline\Lambda$ (where $\overline\Lambda=m_D-m_c$),
is the Borel parameter characterizing the average virtuality
of the c-quark in the correlator, 
$s_0^D$ is the quark-hadron duality threshold ,
and $\mu_c$ $\sim M$ is the factorization scale.
The fact that $L_{n\geq 3}\sim (\Lambda_{QCD})^n$
and  $S_{nk}\sim (1/M)^n$ allows one to retain 
a finite numbers of terms in the sum over $n$ (
$n_{max}=6$ is already providing a sufficient accuracy).

The $f_D\to f_B$ transition in the sum rule (\ref{fDsr})
is realized by replacing 
\begin{equation}
m_c\to m_b,~ m_D\to m_B,~
s_0^D\to s_0^B,~\mu_c\to \mu_b 
\label{replace}
\end{equation}
with the scale-dependence given by the relevant 
renormalization-group factors.
The $c\leftrightarrow b$ correspondence does not mean however that both 
sum rules for $f_B$ and $f_D$ are equally accurate. 
The numerical hierarchies of corrections
are different in the appropriate ranges of Borel parameters,
so that generally the $D$-meson sum rules are less stable.
Nevertheless,  
since $f_B$ is of a crucial importance for 
$B$-physics it is very useful to check the method
by comparing the sum rule prediction for $f_D$
with experiment. 

In Table~2 two recent sum rule determinations
of $f_D$ are presented (for a more detailed overview including 
older results see 
\cite{CK}). Comparison with the lattice QCD results
reveals an encouraging agreement.  
In \cite{KRWWY00} the perturbative part 
is taken into account in $O(\alpha_s)$,  
and the quoted theoretical uncertainty
is largely determined by the $c$-quark mass interval. 
The one-loop pole mass $m_c^{1loop}=1.3\pm 0.1$ GeV was taken, 
which overlaps with the lower part of the $m_c^{1loop}$
interval obtained from the values of $\bar{m}_c(\bar{m}_c)$ 
given in Table 1. One possibility to improve the  sum rule 
is to use the recently obtained $O(\alpha_s^2)$ results
for the heavy-light correlator \cite{Chetyrkin2001},
providing the coefficient $S_{02}$ in Eq.~(\ref{fDsr}). 
This was done in \cite{PeninStein} in the framework of HQET. 
In full QCD so far  only $f_B$ was recalculated \cite{JaminLange} 
with the $O(\alpha_s^2)$ accuracy.

In addition, to have more confidence in the power expansion
of the correlator, it would be useful to calculate the $d=7$ correction 
proportional to a combination of quark and gluon condensates.
A better determination of $m_c$, inclusion of $O(\alpha_s)^2$ 
corrections in full QCD, together with a systematical 
use of $\bar{m}_c(\bar{m}_c)$, are the remaining  resources
of improvement for the sum rule result for $f_D$. 
More difficult is to assess
the ``systematic'' uncertainty related to the quark-hadron duality
approximation in the $D$ meson channel.
\begin{center}
\begin{table}[h]
\vspace{-1cm}
\caption{}
\begin{tabular}{| c | c |c|  }
\hline $f_D$ (MeV) & Ref. & Method \\ 
\hline
&&\\
200(20) & \cite{KRWWY00}& SVZ, $O(\alpha_s)$\\
&&\\
195(20) & \cite{PeninStein}& SVZ+HQET, $O(\alpha_s^2)$ \\
&&\\
\hline
&&\\
203(14)& \cite{Ryan01} & Lattice QCD average  (quench.) \\
226(15)& \cite{Ryan01} & \hspace{0.8cm} ``  \hspace{1.5cm} (unquench.) \\
& &\\
\hline
\end{tabular}
\end{table}
\end{center}
\vspace{-1cm}
Having obtained a prediction for $f_D$ one is not 
yet able to compare it with an experimental number,
because  only $f_{D_s}$ is measured,
the latest result \cite{ALEPH} being :
\begin{equation}
f_{D_s} = 285\pm 19\pm 40~\mbox {MeV}. 
\label{fDsexp}
\end{equation} 
Importantly, QCD sum rules also predict the $f_{D_s}/f_D$ ratio
in terms of $m_s$ and $\langle\bar{s}s\rangle/\langle\bar{q}q\rangle$,
$(q=u,d)$ the ratio of strange and nonstrange quark condensates. 
The rather old results  collected in \cite{CK} yield an interval: 
\begin{equation}
f_{D_s}/f_D= 1.11\div 1.27 \,,
\label{interv}
\end{equation}
to be compared with the recent 
averages 
$
f_{D_s}/f_D= 1.12(2)~[1.12(4)]
$
of lattice quenched [unquenched] QCD \cite{Ryan01} 
with a smaller uncertainty (see also \cite{Rolf}).  
Multiplying the sum rule prediction 
from Table 2: $f_D=200 \pm 20$  MeV 
by the ratio (\ref{interv})  we obtain:
$
f_{D_s} = 240 \pm 40 ~\mbox {MeV}, 
$
in the ballpark
of the experimental interval (\ref{fDsexp}). Improving 
the latter and measuring $f_D$ will provide  
more decisive checks.

\section{$D\to \pi,K$ form factors from LCSR}

Measuring the semileptonic $D\to \pi l\nu_l$ 
decay distribution 
\begin{eqnarray}
\frac{d\Gamma^{(D\to \pi l \bar{\nu})}}{dq^2}=
\frac{G^2|V_{cd}|^2(E_\pi^2-m_\pi^2)^{3/2}}{24\pi^3}
[f^+_{D\pi}(q^2)]^2
\nonumber
\\
+ O(m_l^2)\,,
\label{dGamma}
\end{eqnarray}
at $m_l^2<q^2<(m_D-m_\pi)^2$ 
and dividing out $|V_{cd}|$ one is able to 
reproduce the experimental values of the $D\to \pi$ form factor
$f^+_{D\pi}(q^2)$. With a sufficiently large statistics,
of semileptonic decays with $l=\mu$, 
the scalar form factor $f^0_{D\pi}(q^2)$ entering the chirally 
suppressed $O(m_l^2)$ part of the decay distribution 
can also be extracted. The Cabibbo enhanced $D\to K l \nu_l $ decays  
provide $D\to K$ form factors with an even better accuracy.

Heavy-to-light form factors are calculated 
using QCD light-cone sum rules (LCSR). 
One of the first applications of this sum rule technique 
was the calculation of $f^+_{D\pi,DK}$ in \cite{BBD}.
The updated results for $f^+_{D\pi,DK}$ including higher twist
terms \cite{BBKR} 
and $O(\alpha_s)$ corrections \cite{Bpialphas} can be found 
in \cite{KRWWY00}. The form factor $f^0_{D\pi}$ 
was calculated in \cite{KRW}.

The sum rules for $B\to \pi$ are obtained
by the same replacement (\ref{replace}). 
Let me emphasize that this transition
is done from one finite mass to the other. Contrary to HQET relations
between $B$ and $D$ form factors, no  
heavy-quark mass approximation is involved.  
Since $f^+_{B\pi}(q^2)$ is used to extract $|V_{ub}|$
from $B\to \pi l \nu$, a comparison 
of the sum rule predictions for $D\to \pi$ form factors with 
experimental data will ensure more confidence in the LCSR method.

The most recent LCSR result 
$f_{D\pi}(0)= 0.65 \pm 0.11 $ obtained in \cite{KRWWY00} 
takes into account the twist-2 term in NLO and twist-3,4 contributions 
in LO (in the expansion of the underlying vacuum-pion
correlator in the  pion distribution amplitudes with growing twist).
From the same correlator, using double dispersion relation
one has access to the $D^*D\pi$ coupling $g_{D^*D\pi}$ 
\cite{BBKR,KRWYcoupl},
predicting the product $f_{D^*}f_Dg_{D^*D\pi}$.
Using the SVZ sum rule result for $f_D$ quoted above, 
the $D^*$-pole contribution
to the $D\to \pi$ form factor is calculated.    
The two  predictions of LCSR are used \cite{KRWWY00} to fit 
the $D\to \pi$  form factor 
in the whole kinematical region $0<q^2<(m_D-m_\pi)^2$
to the simple ansatz \cite{BecKaid}
inspired by dispersion relation:  
\begin{equation}
f_{D\pi}^+= \frac{065\pm 0.11}{(1- q^2/m_{D^*}^{2})
(1-\alpha^{D\pi}q^2/m_{D^*}^2)},
\label{fDpi}
\end{equation}
with $\alpha^{D\pi}= 0.01 ^{+0.11}_{-.07}$.
Interestingly, the sum rule results are consistent 
with the $D^*$-pole dominance for the form factor.
The predicted integrated decay width 
$\Gamma(D^0\rightarrow \pi ^- l^+ \nu_l)/|V_{cd}|^2 = 
0.13 \pm 0.05 \quad \mbox{ps}^{-1}~$
is in a reasonable agreement with the experimental number \cite{PDG02}
$\Gamma(D^0\rightarrow\pi^- e^+ \nu_e)/|V_{cd}|^2 =
0.174 \pm 0.032\quad \mbox{ps}^{-1}$.
At the zero momentum transfer the LCSR prediction also agrees with 
the recent lattice result \cite{APE}  
$f_{D\pi}(0)=0.57(6)\pm^{0.01}_{0.00}$. 
On the other hand, the lattice calculation 
\cite{APE} suggests that the contribution of excited $D^*$ states 
is not small. 
To assess the reliability of these theoretical
predictions, one has to wait until the $D\to \pi l \nu_l$  
decay distribution is accurately measured.

Finally, the LCSR result 
for $D\to K$ form factor \cite{KRWWY00} 
is $f_{D K}(0)=0.78\pm 0.11$ 
at $m_s(1 \mbox{GeV})=150$ MeV, very sensitive to the strange quark
mass. The corresponding integrated width 
$
\Gamma(D^0\rightarrow K ^- l^+ \nu_l)/|V_{cs}|^2  = 
0.094 \pm 0.036\quad \mbox{ps}^{-1},
$
 is in a good agreement 
with experiment \cite{PDG02} 
$
\Gamma(D^0\rightarrow K^- l^+ \nu_l)/|V_{cs}|^2 =
0.087 \pm 0.004 \quad \mbox{ps}^{-1}$. 

Concluding this section, let me briefly discuss the  
important issue of the $D^*D\pi$ coupling (see also \cite{MF}).
The recent first measurement of the 
total width of $D^*$ by CLEO
collaboration \cite{CLEODstar}: 
$\Gamma_{tot}(D^*)= 96 \pm 4 \pm 22$ keV yields 
for this coupling 
$g_{D^*D\pi} = 17.9 \pm 0.3 \pm 1.9$ (using the definition 
of Ref.~\cite{BBKR}).
The LCSR prediction \cite{BBKR,KRWYcoupl}, 
$g_{D^*D\pi} = 10 \pm 3.5 $, is obtained by dividing the 
calculated product $f_Df_{D^*}g_{D^*D\pi}$ by the two decay constants,
$f_D$ and $f_{D^*}$, extracted from 2-point SVZ sum rules. 
Taking into account the estimated theoretical uncertainty,
the  upper limit of the LCSR prediction is $g_{D^*D\pi} = 13.5 $,
still  25\% lower than the central value of the CLEO number. 
Meanwhile, the first lattice QCD prediction 
$g_{D^*D\pi} = 18.8\pm 2.3_{-2.0}^{+1.1}$
\cite{latticCoupl} agrees with the CLEO result. If the future measurements 
(although extremely difficult !) 
and lattice  calculations confirm the large value
of this coupling, one has to clarify 
the status of the LCSR prediction.
Having in mind that all other sum rules discussed above 
use  one-variable dispersion relations,
one might suspect that certain complications arise in 
the double dispersion relation used in  LCSR for
the $D^*D\pi$ coupling. More specifically,  the simplest quark-hadron
duality ansatz (one resonance
plus continuum) in both $D$ and $D^*$ channels 
may be too crude. One possible scenario 
was recently discussed in \cite{BCLYOPP}:     
assuming a partial cancellation between the contributions of 
excited and ground $D,D^*$ states in the dispersion relation,
one gets an increase of the LCSR coupling. 
Without going into further details, 
let me only mention that the magnitude of the coupling $g_{D^*D\pi}$ 
and the shape of the form factor $f^+_{D\pi}(q^2)$ 
are closely related. Suppose the form factor 
is dominated by the $D^*$-pole contribution:
\begin{equation} 
f^+_{D\pi}(q^2) =\frac{f_{D^*}g_{D^*D\pi}}{2m_{D^*}(1-q^2/m_{D^*}^2)}.
\end{equation}
Taking  for $g_{D^*D\pi}$ the CLEO central value 
and multiplying it with $f_{D^*}=250$ MeV
(within the lattice QCD prediction \cite{fDstarlatt}) 
one obtains a semileptonic width
$\Gamma(D^0\rightarrow \pi ^- l^+ \nu_l)/|V_{cd}|^2 = 
0.37  \quad \mbox{ps}^{-1}~$  which is about two times larger 
than the experimental width quoted above \cite{PDG02}.
To make the strong coupling measured by CLEO 
consistent with the semileptonic width, one needs 
a substantial negative interference 
between the $D^*$-pole  and  excited $D^*$ states
in the dispersion relation for the form factor 
(as also noticed in \cite{BCLYOPP}), resulting
in a visible deviation of the $D\to \pi l \nu_l$ decay distribution
from the $D^*$-pole dominance.

\section{Rare $D$ decays}

Exclusive rare $D$ decays such as $D\to \mu^+\mu^-,2\gamma$, etc. 
will become important highlights in the future high-statistics 
charm physics experiments. Being suppressed in the Standard Model,
these decays  are promising indicators 
of new physics. The long-distance amplitudes of rare $D$ decays
in the Standard Model still lack a QCD based analysis. 
I believe, QCD sum rules in both two-pont (SVZ) and light-cone versions
can essentially help in solving this problem, 
One example is 
the LCSR prediction for weak radiative decays obtained in \cite{KSW96}:
\begin{eqnarray}
&BR(D^{+} \to \rho^{+} \gamma)= 2.7 \cdot 10^{-6}\,,\nonumber\\
&BR(D^0 \to \rho^0 \gamma)= 3.0\cdot 10^{-6}\,, \nonumber\\
&BR(D_s \to \rho^+ \gamma)= 2.8\cdot 10^{-5}\,, 
\end{eqnarray}
where all numbers have an $O(50\%)$ accuracy. 
This analysis can be further improved and extended
to the other rare $D$ decay channels.

\section*{Acknowledgments}
I am grateful to the conveners and organizers 
for a  very useful and informative workshop.
This work is supported by the German Ministry 
for Education and Research (BMBF).

\end{document}